\documentclass[a4paper]{jpconf}
\usepackage{graphicx}
\begin{document}
\title{The dynamics of the radiative zone of the Sun}
\author{S Turck-Chi\`eze$^1$, S Couvidat$^2$, V Duez$^3$, S Mathis$^1$, J Marques$^4$, A\,\,Palacios$^5$, L Piau$^1$}
\address{$^1$IRFU/ CEA/ CE Saclay, Gif sur Yvette, France $^2$ HEPL, Stanford, USA $^3$Argelander-Institut fur Astronomie,  Universitat Bonn, Germany $^4$ Observatoire de Meudon, Meudon, France $^5$GRAAL Montpellier, France}
\ead{sylvaine.turck-chieze@cea.fr}
\begin{abstract}
Helioseismology puts strong constraints on the internal sound speed and on the rotation profile in the radiative zone. Young stars of solar type are more active and faster rotators than the Sun. So we begin to build models which include different rotation histories and compare the results with all the solar observations. The profiles of the rotation we get have interesting consequence for the introduction of magnetic field in the radiative zone. We discuss also the impact of mass loss deduced from measured flux of young stars. We deduce from these comparisons some quantitative effect of the dynamical processes (rotation, magnetic field and mass loss) of these early stages on the present sound speed and density. 
We show finally how we can improve our present knowledge of the radiative zone with PICARD and GOLFNG.\end{abstract}
\section{Time evolution of the internal rotation induced by its transport of momentum}
The Sun is rotating and the internal seismic rotation pushes us to introduce the effect of the time evolution of the rotation in solar model. Turck-Chi\`eze et al., 2010a have followed three rotation histories: a very low initial rotation, an academic case where the initial value is so low that the Sun naturally slows down, and two higher initial rotations which lead to 20 and 50 km/s at the arrival on the main sequence plus some braking to reach the present surface value of 2.2 km/s. Their time internal evolution  are followed, modified by the transport of momentum induced in the radiative zone by both advection and diffusion terms.

We notice that a radial differential rotation is present in all the models we have computed and that its radial profile is mainly established during the contraction phase which corresponds to the first million years (see Figure 7 of Turck-Chi\`eze et al. 2010a and tables 4 and 5). It seems that a gradient stays present in the core of the Sun today if one believes the detection of gravity modes with the GOLF instrument aboard SoHO (Turck-Chi\`eze et al. 2004, Garcia et al. 2007, Mathur et al.2008).
Of course in our computations, the order of magnitude of the gradient depends strongly on the presence or not of braking. Our  conclusions on the transport of momentum by rotation alone are the following:
\begin{figure}
\begin{center}
\includegraphics[width=16pc]{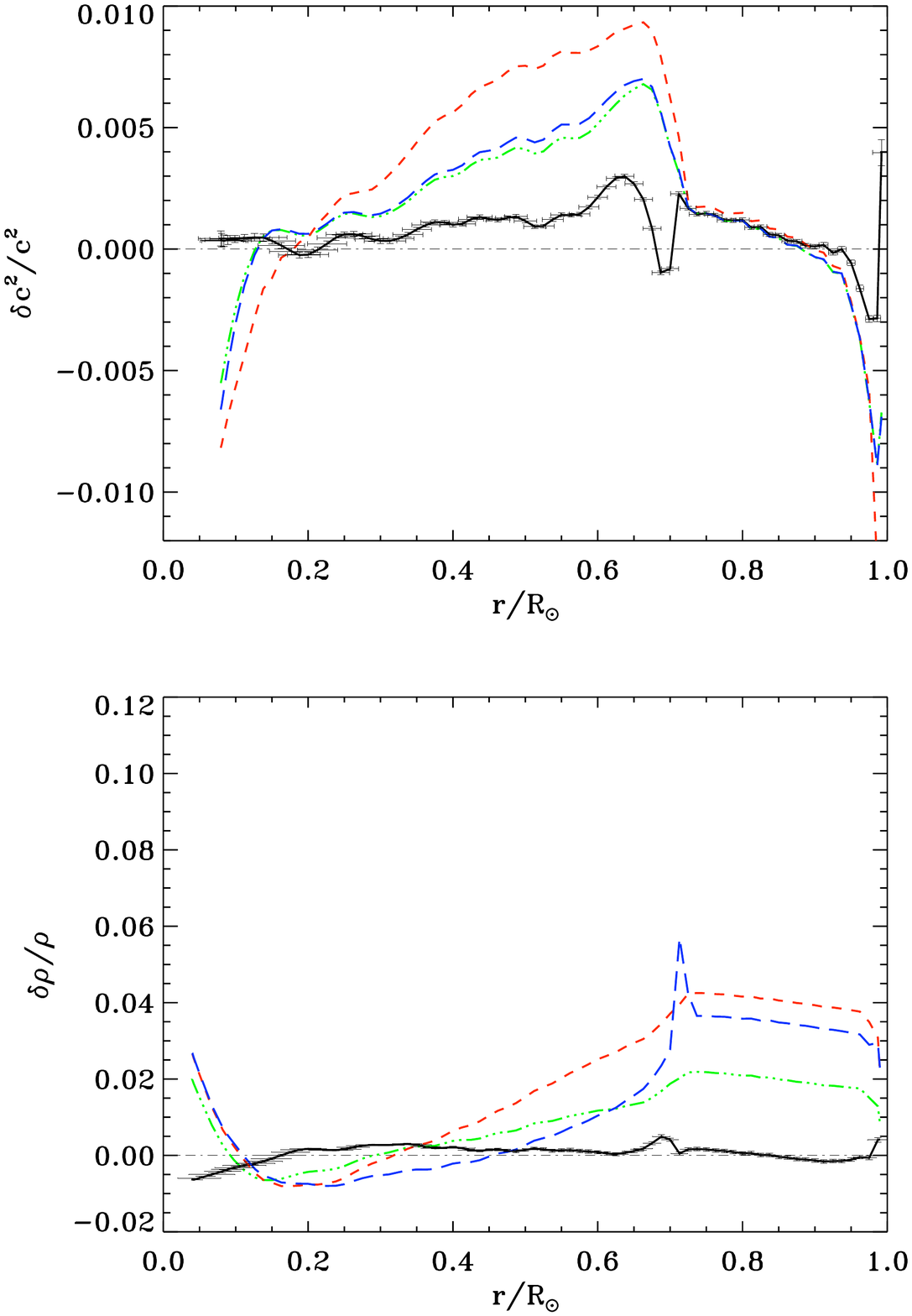}
\includegraphics[width=16pc]{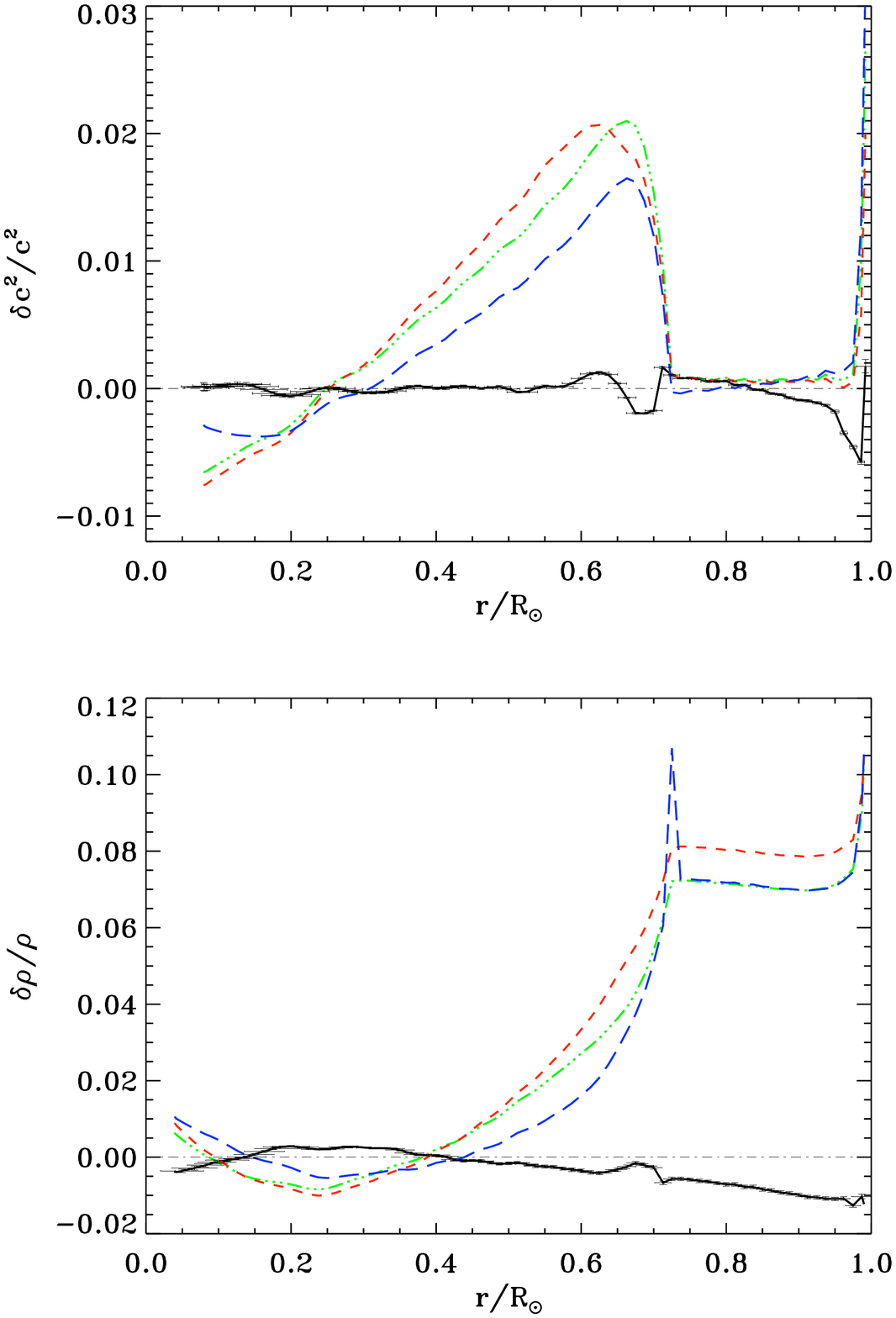}
\end{center}
\vspace{-0.5cm}
\caption{\label{trans}Left: 
Radial squared sound speed and density differences between
observations and models. Seismic model is in black with error bars coming from the
seismic data.  The standard GN composition model, model A  (slow rotation) and model B (large rotation) are respectively in green (dotted dashed line),
blue (large dashed line) and red (small dashed line). Right: Idem: now the standard model calculated with the most recent Asplund composition, model with turbulence in the tachocline  and model with 0.33 M$_\odot$ are respectively in red (dashed line),
green (dot dashed line) and blue (large dashed line). From Turck-Chi\`eze et al. 2010a,b.}
\end{figure} 

1) This transport  during the main sequence appears extremely small.  The meridional circulation in the radiative zone is smaller by about 10 orders of magnitude in comparison with the observed convective meridional circulation velocity measured at 99 \% R$_\odot$. 

2) The sound speed and density profiles are not largely modified by rotation alone but the difference with obervation slightly increased (Figure1 left). 

3) Although the combined effect of meridional circulation and shear-induced turbulent associated to rotation is small, this study leads to radial rotation profiles that can be directly compared to the seismically observed one. The comparison with observations sustains the idea that the Sun was not a rapid rotator (Figure 2 left).  The low initial rotation (the academic case) model helps us to quantify the structural consequences of rotation.

4) We believe that the radial rotation profile during the contraction phase could be reduced by the transport of momentum by magnetic field in the core and that the diffusion of this field can flatten the profile in the rest of the radiative zone. A first tentative has been performed by Eggenberger et al. (2007), but the stability of that field, if one believes that it is still present, supposes probably a mixture of poloidal and toroidal components.  

It is why we have begun to introduce a magnetic configuration which may exist in the solar-type radiative regions and developed the modified structural equations.  Probably, these fields cannot exceed several MG. If it was not the case, their impact on the solar shape would have been clearly established already. Consequently,  the $\beta$ parameter that gives the ratio between the total pressure and the magnetic pressure stays large in the whole radiative zone and the microscopic quantities including the sound speed must not be modified significantly by the presence of such a field. In Duez et al., 2010, we have calculated the solar quadrupole moment which results from such a field. So the order of magnitude of the variation of the different terms at the arrival on the main sequence is small (Duez, Mathis \& Turck-Chi\`eze, 2010). But the description of the early stage is more complex and justifies to look deeper on young stars.

\section{The strong activity of the young solar like stars}
Young solar-like stars are generally more active and are rotating   faster than the present Sun. The review of Gudel (2007) gives some specific properties of  young analogs of the Sun. Their observation confirms the need  to introduce  magnetic field in stellar modelling to derive   an appropriate comparison with all the present observables.

The specific observation of  young stellar rotations shows different  temporal rotation profiles during the contraction phase and the PMS (Figures 1 and 3  of Bouvier 2008a). If a typical braking law ${dJ} /{dt} \propto \Omega^3$ is observed between $\alpha$ Persei, Pleiades, Hyades and the Sun, other observations of young stars show that this law is not universal and that it exists slow and fast young rotators. This difference may be explained by the coupling timescale of core and envelope (Bouvier 2008b).

The magnetized stellar winds are efficient to brake the magnetically active stars. Their efficiency is about a factor 1000 greater than any other phenomenon. So, even their effect in the radiative zone appears small for the present Sun,  magnetic field of the young Sun is the central ingredient which governs the rotational evolution of solar-like stars and probably already in the very early stage. We have not yet results for this phase because one needs to know how to introduce a potential growing of the field probably through a dynamo process. We begin to introduce another effect connected to UV loss observed in young analogs, in modelling a   mass loss in the early stage directly deduced from the study of a lot of young stars. 

\section { Young Sun mass loss and its impact on  luminosity and its present sound speed}
In standard model studies, the lost of luminosity by X and UV due to the  strong activity of the young Sun is ignored. For the early stage, these phenomena are measured in young stars (Ribas, 2009). Moreover we now know  that the present dynamo, which produces the Hale cycle, leads to luminosity loss 10$^6$ greater than was supposed by the standard model along the last decade (about some 10$^{-3}$ instead of 10$^{-9}$).  So it is clear that one needs  to introduce the lost of luminosity along age produced by different phenomena.

To progress on this direction, we have used the expressions given by Ribas to estimate some properties of the Sun in the first stage, knowing that there is  difficulty to separate periods of accretion and mass loss and the duration of each of them. Our first model leads to a more luminous Sun  during the formation of planets phase which can help to resolve partly the solar paradox. Figure 1 right illustrates the impact of mass loss on  the sound speed and density. One confirms that contrary to the effect of transport of momentum by rotation (figure 1 left),  the introduction of mass loss in the early stage  could reduce the  discrepancy that we observe with the observed sound speed and density at the present time.

\begin{figure}
\begin{center}
\includegraphics[width=14pc]{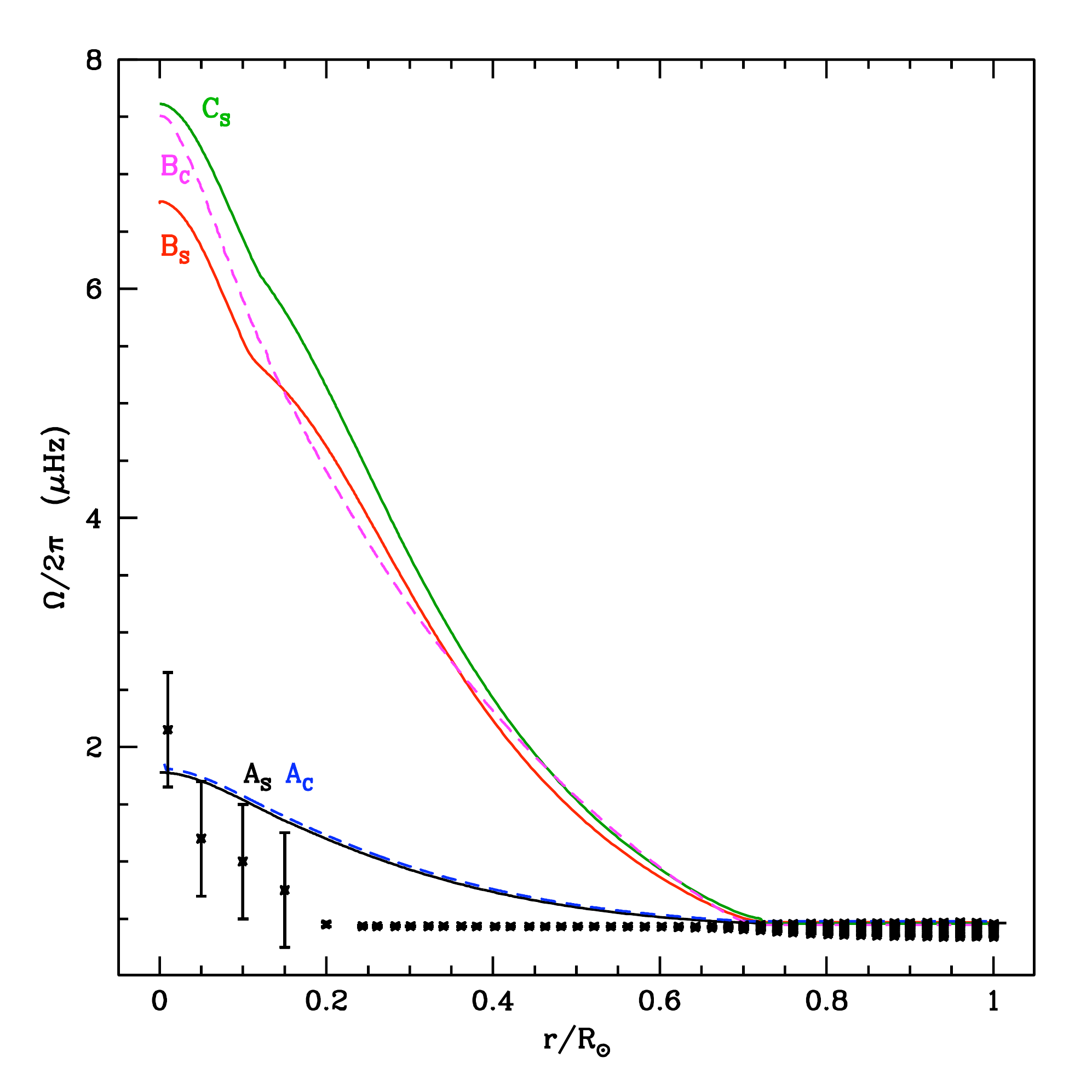}
\includegraphics[width=22pc]{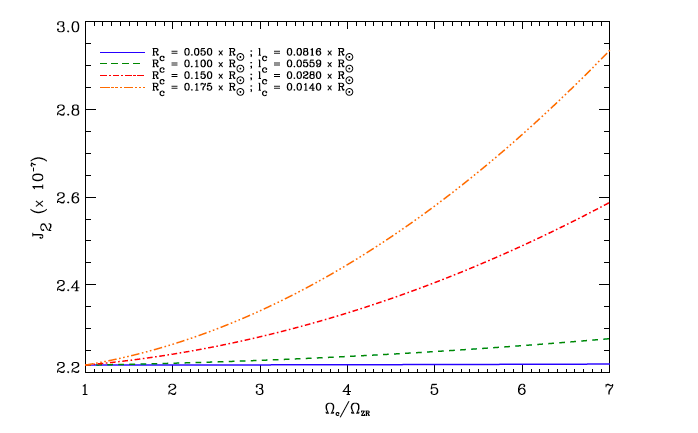}
\end{center}
\vspace{-0.5cm}
\caption{\label{trans}Left: Comparison between the solar internal profile predicted by
    different models (A for slow initial rotation, B and C for fast initial rotation) and that deduced from helioseismology. The data down to r/R$_\odot$ = 0.2 are deduced  from the acoustic mode splittings determined
by GOLF, MDI and GONG instruments (Couvidat et al. 2003b, Eff-Darwich,
2008). The data in the core mimic the information extracted from the  gravity mode study. From Turck-Chi\`eze et al. 2010a). Right: Quadrupole moment for different profiles of rotation versus the ratio central rotation over surface rotation. From Duez, Turck-Chi\`eze, Mathis 2011.}
\end{figure} 

\section{The need for new observational constraints}
The whole story of the solar activity supposes to introduce one by one all the different processes which maintain this activity up to now.  To check its relevance,  all the observations we may gather on young cluster stars or on the present Sun are useful. It is why it will be particularly useful to predict how internal magnetic field and rotation profile impact on the shape of the present Sun and how this shape evolves at small and large time scales in using the measurements of SODISM/PICARD satellite (Thuillier, Dewitte and Schmutz 2007).  We show,  in Figure 2 right, how different profiles of rotation (a flat profile outside the core and an increase below R$_C$) could influence the solar quadrupole moment. We find that they are of the same order than the effect of a deep field. Up to now, the aspheric shape of the Sun is not yet understood,
the balloon shows values of oblateness of 4.3 to 10. 10$^{-6}$ increasing with the cycle (Emilio
et al., 2007) and we can hope to determine the different components of this shape with PICARD.

In addition to SODISM, a better description of the solar core after the SoHO measurements is of fundamental  importance to know precisely the density, temperature and rotation profiles.  It is for such objective, that we have prepared the successor of GOLF/SoHO, through the building of the GOLF-NG prototype for which we have now demonstrated all the improved performances and solved all the technical difficulties . All the tests  done have successfully shown  the ability of this instrument to measure properly the time variation of the Doppler velocity on at  least 6 heights in the atmosphere in order to eliminate part of the granulation noise and to measure quicker the small amplitude of useful gravity modes (Turck-Chi\`eze et al. 2006, 2008, 2011).
\section*{References}
\begin{thereferences}
\item Bouvier J 2008a in {\it Stellar Magnetism}, EDP science, p 199
\item Bouvier J 2008b {\it A\&A}  {\bf 489} L53
\item Couvidat S, Turck-Chi\`eze S \& Kosovichev A 2003a {\it ApJ} {\bf 599} 1434
\item Couvidat S, et al. 2003b {\it ApJ} {\bf 597} L77 
\item Duez V, Mathis S \& Turck-Chi\`eze S 2010 {\it MNRAS} {\bf 402} 271
\item Duez V, Turck-Chi\`eze S \& Mathis S 2011 {\it A\&A}  in preparation
\item Emilio M. et al. 2007  {\it Astrophys. J.} 660L 161
\item Garcia R A, Turck-Chi\`eze S et al. 2007 {\it Science} {\bf 316} 1537
\item Garcia R A et al. 2008 {\it Astron Notes} {\bf 329} 476
\item Eff Darwich A et al. 2008 {\it ApJ} {\bf 679} 1636
\item Eggenberger P, Maeder A \& Meynet G 2005 {\it A\&A} {\bf 440} L9
\item Gudel M 2007 {\it Living Review Solar Phys} {\bf 4 } 3
\item Mathis S \& Zahn JP 2004, {\it A\&A}, {\bf 425} 229
\item Mathur S, Eff-Darwich A, Garcia R A\& Turck-Chièze S. 2008 {\it A\&A} {\bf484} 517
\item Thuillier G, Dewitte S \& Schmutz W 2007 {\it Adv. Space Res. } {\bf 26} 1792
\item Turck-Chi\`eze S. et al. 2001 {\it ApJ} {\bf 555} L69 
\item Turck-Chi\`eze S et al. 2004a  {\it Phys Rev L}  {\bf 93} 211102
\item Turck-Chi\`eze et al. 2004b {\it ApJ} {\bf 604} 555
\item Turck-Chi\`eze S. et al. 2006 {\it Adv. Space. Res.} {\bf 38} 1812
\item Turck-Chi\`eze S. et al. 2008 {\it Astron Notes} {\bf 329} 521
\item Turck-Chi\`eze S. et al. 2010a {\it ApJ} {\bf 715} 1539
\item Turck-Chi\`eze S. et al. 2010b {\it ApJ lett} submitted
\item Turck-Chi\`eze S. et al. 2011 {\it Experimental Astronomy} in preparation
\end{thereferences}
\end{document}